\pdfoutput=1
\documentclass[doublecol]{epl2}
\usepackage{graphicx}
\usepackage{amsmath}
\usepackage{times}
\usepackage{amssymb}
\usepackage{mathrsfs}
\usepackage{color}
\usepackage{hyperref}
\usepackage{version}

\usepackage[utf8]{inputenc}
\usepackage[T1]{fontenc}

\newcommand{\kb}{k_\text{\tiny B}}

\newcommand{\fig}{FIG.}

\providecommand{\avg}[1]{\left \langle #1 \right \rangle}

\newcommand{\beq}{\begin{equation}} \newcommand{\eeq}{\end{equation}}

\title{Probing active forces via a fluctuation--dissipation relation:\\
  application to living cells}

\author{P. Bohec\inst{1} \and F. Gallet\inst{1} \and C. Maes\inst{2}
\and S. Safaverdi\inst{2} \and
P. Visco\inst{1}\thanks{Correspondig Author:
\email{paolo.visco@univ-paris-diderot.fr}} \and F. van Wijland\inst{1}
}
\shortauthor{P. Bohec \etal}
\institute{
\inst{1} Laboratoire Mati\`ere et Syst\`emes Complexes, CNRS UMR
  7057, Universit\'e Paris Diderot --  10 rue Alice Domon et L\'eonie
  Duquet, 75205 Paris cedex 13, France \\
\inst{2}  Instituut voor Theoretische Fysica, KU Leuven --
Celestijnenlaan 200D,
 B-3001 Leuven, Belgium}

\date{\today}

\abstract{
We derive a new fluctuation--dissipation relation for non--equilibrium
systems with long term memory. We show how this relation allows one to
access new experimental information regarding active forces in living
cells that cannot otherwise be accessed. For a silica bead attached to
the wall of a living cell, we identify a cross-over time between
thermally controlled fluctuations and those produced by the active
forces. We show that the probe position is eventually slaved to the
underlying random drive produced by the so-called active forces.}

\begin{document}

\maketitle

Living cells are paradigmatic out of equilibrium systems, subjected to
the ATP-driven activity of a collection of molecular motors, whose
individual motion cannot easily be disentangled from thermal
fluctuations. These are relatively small systems for which fluctuation
phenomena are prominent, as investigated in recent
works~\cite{lau,bursac2005,lenormand,PhysRevLett.100.118104,PhysRevLett.103.090601,gallet09,fletcher09,robert,gov2011},
with specific focus on the active forces and non-Newtonian rheology
~\cite{lau,bursac2005,lenormand, mizuno2007} which can be measured via
micro-rheological devices. Our goal in the present work is to show how
the cell body's random pull-and-push can be investigated by means of
very recent theoretical advances in the field of non--equilibrium
statistical mechanics. And indeed, from the theory standpoint, in the
recent past, much effort has been invested in deriving simple
generalizations or extensions of the celebrated
fluctuation--dissipation theorem for systems that can be arbitrarily
far out of
equilibrium~\cite{falcioni90,falcioni95,speck06,marconi08,villamaina08,villamaina09,chetrite08,speck09,baiesi09a,baiesi09b,baiesi11,baiesi10,PhysRevLett.103.090601,seifert10}.
These
efforts have given birth to a flurry of formulas relating the response
of a system to a small external perturbation to some correlation
functions, even when the system is not in an equilibrium state. Yet,
so far none of these formulas has been of any predictive power in an
actual experimental system. Existing experiments revolving around
these theoretical advances have been confined to refined and
nontrivial confirmations that in some small scale systems such as an
optically trapped Brownian
particle~\cite{blickle06,blickle07,solano09} the various ingredients
entering these extended fluctuation-dissipation relations (EFDR) can
indeed be measured. Given that the dynamics of living cells exhibit
strong memory effects, we will first have to derive our own version of
an EFDR adapted to a system with stochastic yet non-Markovian
dynamics. The latter EFDR and its consequence for the understanding of
active forces are the central topics of this letter. For instance, how
relevant is the long term memory in relating response and
fluctuations? What new pieces of information on non--equilibrium
forces can be learnt from such a relationship?  Can thermal
fluctuations be disentangled from those arising from the active forces
in a quantitative way? These are the questions we wish to address in
the present work.

\begin{figure}
\begin{center}
\includegraphics[width=\columnwidth]{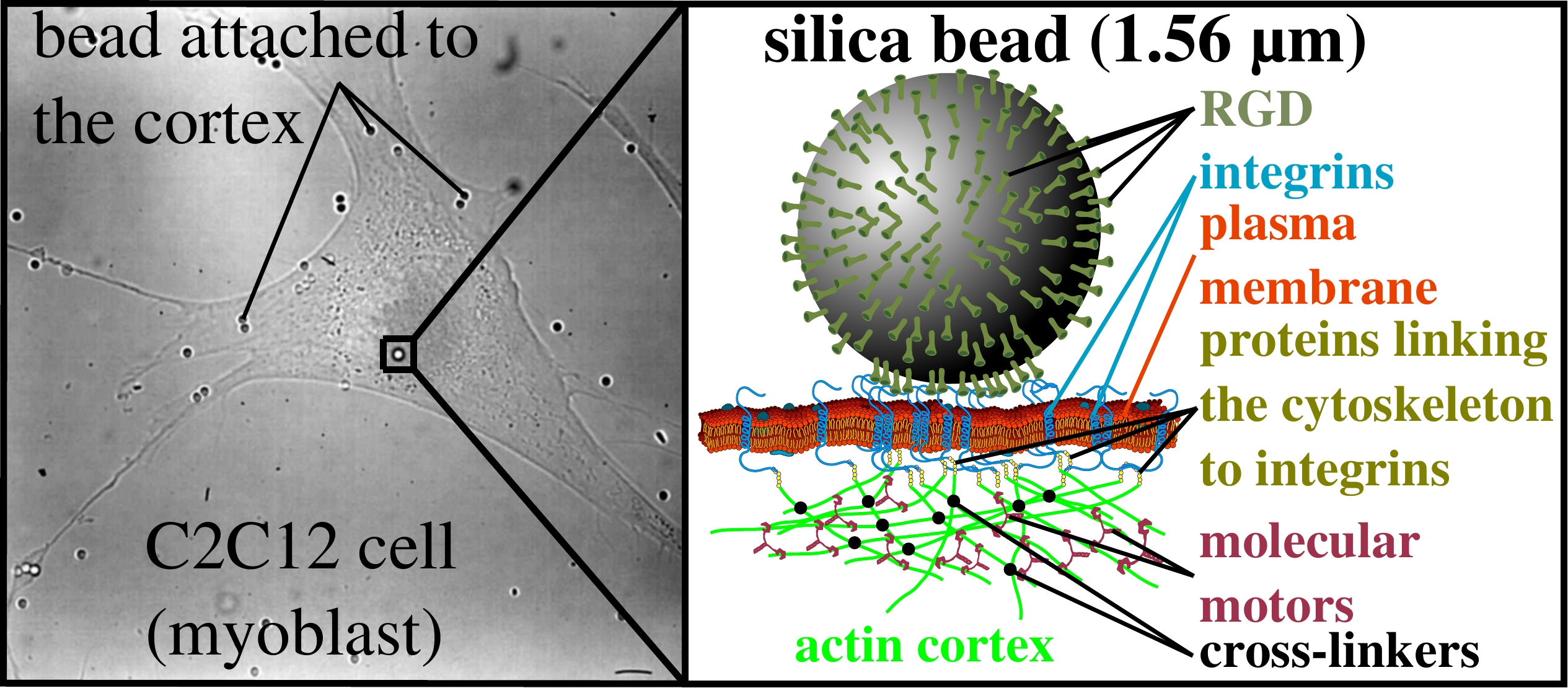}
\end{center}
\caption{Schematic view of the bead probe as attached to the
  cytoskeleton. The active forces are the by-product of the ATP-driven
  molecular motors activity.
   \label{fig:setup}}
\end{figure}

Our experimental system, which has already been considered
in \cite{gallet09} for other purposes, consists of a pre--muscular
(C2C12) cell to which we slave a micron sized silica probe
specifically bound to the membrane (in \cite{bursac2005} the authors
used a human airway smooth muscle cell). Its dynamics strongly feels
the visco--elastic non-Newtonian underlying medium~\cite{bursac2005},
which gives rise to memory (non-Markovian) effects. 
A schematic view of our experimental setup is shown in
\fig~\ref{fig:setup}.
According to \cite{lau,hoffman05,lenormand04,desprat05,wirtz09}, we
consider that linear-viscoelasticity correctly describes the
mechanical cell behavior. The bead thus evolves under the combined
effect of thermal fluctuations $\xi(t)$ and active forces $F_a(t)$,
which represent all the non-thermal stochastic effects due to cellular
activity.  Discarding inertial effects, the position $x$ of the bead
is governed by the Langevin equation
\begin{equation}
\label{eq:langevin}
\int_{-\infty}^t \! dt' \, \gamma(t-t') \frac{d x(t')}{dt'} = F_a(t)
+ \xi(t)\,\,,
\end{equation}
where $\xi$ a Gaussian colored noise with correlations
$\sigma(\tau)=\langle{\xi(t) \xi(t+\tau)}\rangle$. The system is
thermalized at an inverse temperature $\beta=1/\kb T$, and we assume
that active forces will not affect the thermal fluctuations, which
then verify a local equilibrium condition
$\sigma(\tau)= \gamma(|\tau|) / \beta$. This assumption is discussed
in~\cite{harada2006} and relies on a time-scale separation which is
experimentally observed in our system, as we shall show later. The
active forces are modeled by an isotropic stationary random force
$F_a(t)$. While the presence of the memory kernel $\gamma(\tau)$ looks
like an innocuous extension of the well-known overdamped Langevin
equation (for which $\gamma(\tau)$ is to be replaced by a Dirac delta
in time), we shall soon see that a number of complications emerge.\\
We begin with presenting the new theoretical result we wish to
exploit, namely a relationship between three quantities, two of which
are accessible to measurement. Let $R(t)$ be the impulse response
function of the bead position at time $t$. Let also $\Delta
x^2(t)=\langle{(x(t)-x(0))^2}\rangle$ be the mean square displacement
of the bead's position. Our goal is to connect $R$ and $\Delta x^2$ to
the statistics of the active force $F_a$.  \\
We start by considering the effect on the bead of an applied force
$f_{exp}[x(t),t]$, which explicitly depends on time $t$ and possibly
on position $x$. The impulse response of the bead measures the effect
of the applied force on the average position, and is defined by the
convolution $\delta \langle{x}\rangle (t)=\int dt'
R(t-t') \avg{f_{exp}[x(t'),t']}$, where $\delta \avg{x}$ is the change
in average position caused by the applied force. In a viscoelastic
medium this is often described by means of the creep
function~\cite{qian2000}, which is simply the derivative of our
response function~\cite{squires2010}.  At equilibrium the
Fluctuation--Dissipation Theorem relates the response to the mean
square displacement $\Delta x^2(t)$. By generalizing this result, it
can be shown that an extra correlation function $\avg{F_a(t') x(t)}$
is involved (technical details are given in~\cite{soghra}). Since we
are dealing with a non--stationary system which exhibits anomalous
diffusion, we shall only consider the time--translation invariant part
of the correlation function, which is equivalent to look at the
function $\avg{F_a(0) x(t)}$. Its expression is given
by~\cite{soghra}:
\begin{multline}
\label{eq:fax2}
\avg{F_a(0) x(t)} = \frac{\beta}{2} \int_{-\infty}^{t} dt'
\sigma(t-t') \frac{d}{d t'} \Delta x^2(t') \\ - \int dt' R(t')
\sigma(t-t') \,\,.
\end{multline}
 It is worth noting that the last term above corresponds exactly to
  the thermal noise--position correlation function $\avg{\xi(0)
  x(t)}$. When there is no active force, (\ref{eq:fax2}) yields to the
  standard result:
\begin{equation}
R(t)=\frac{\beta}{2}  \frac{d}{dt}  \Delta x^2(t)\,\,,\quad \text{for $t>0$.}
\end{equation}
The originality of our approach is summarized in formula
(\ref{eq:fax2}). This formula expresses the cross-correlation between
two observables, namely the force exerted on the bead and its
position, while most of the published work in the field focuses on the
autocorrelation function of a single observable (force or
position). Indeed, several recent works attempt to relate the force
autocorrelation function to the mean square displacement of the
probe~\cite{bursac2005,robert,gallet09,mizuno09}. Alternatively, the
calculation derived here provides an explicit expression for
$\avg{F_a(0)x(t)}$, and makes this quantity easily computable from
experiment, provided that one is able to measure the mean square
displacement of the bead and its impulse response function, as shown
below.

\begin{figure}
\begin{center}
\includegraphics[width=\columnwidth]{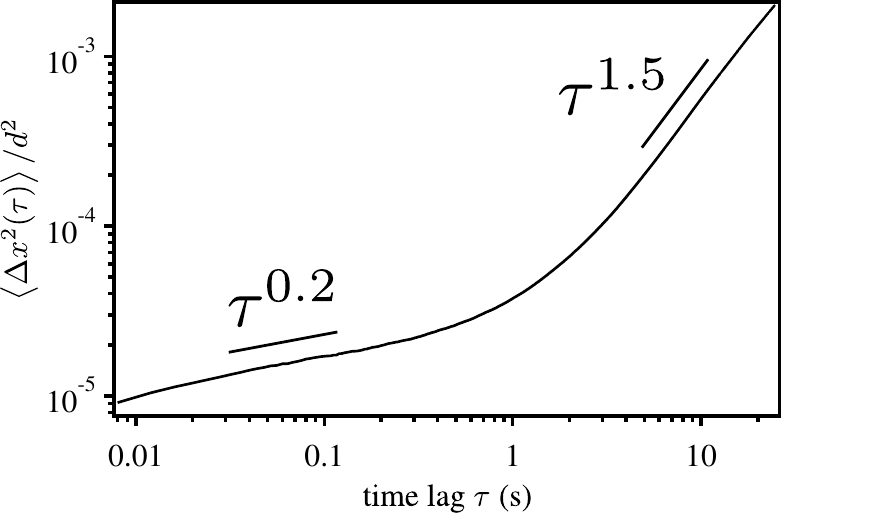}
\end{center}
\caption{The probe mean square displacement as a function of time when
  there is no external applied force. A definite crossover time is
  found at $\tau\simeq 1$ s.  \label{fig:msd}}
\end{figure}

\begin{figure}
\begin{center}
\includegraphics[width=\columnwidth]{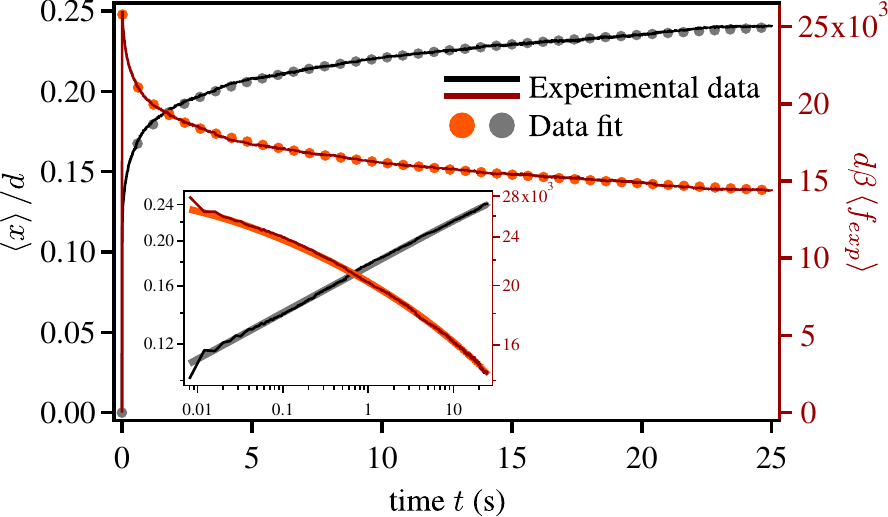}
\end{center}
\caption{Average particle position (black) and applied force (red) in
  dimensionless units ($d$ is the particle diameter), as a function of
  time. Inset shows the same plot on a Log-Log scale. Data are very
  accurately interpolated by a sum of two power
  laws.  \label{fig:posforce}}
\end{figure}

We shall now apply our result to probe active forces in living cells.
The experimental set-up is the same as exposed in~\cite{gallet09} and
consists of a small bead attached through RGD-integrin links to the
cortex of a pre-muscular cell (see \fig \ref{fig:setup}). First, we
tracked the free diffusion of the bead on the cell cortex at room
temperature ($\sim 298 \text{K}$). This allowed us to measure the mean
square displacement of the probe as a function of time $\avg{\Delta
x^2(t)}$, shown in \fig~\ref{fig:msd}. Possible mechanical drifts
obtained from the motion of a bead attached to the glass substrate
have been subtracted for each experiment. The mean-square displacement
exhibits a two-step growth characterized by an early sub-diffusion, up
to a 1s timescale yet to be interpreted, followed by super-diffusion
also observed in~\cite{caspi}. The 1.5 exponent measured
from \fig~\ref{fig:msd} in the super-diffusive regime may vary in the
range 1.3-1.7, according to the density of RGD ligand at the bead
surface. Thus its value has no universal character, but it is close to
the 3/2 prediction one would get assuming a linear decay for the
overall noise terms (i.e.  thermal--noise plus active force
correlation function)~\cite{lau}. It is also worth to note that we do
not see any saturation effects to the mean square displacement, as the
area covered by the bead after 25s is about only 1\% of the projected
bead surface. This corroborates our modelling in terms of a
generalized Langevin equation (\ref{eq:langevin}) for a free
particle. Of course at larger time scales it is possible that some
caging effect from the actin filaments and microtubules gives rise to
an effective confining potential. The effect of this potential on the
mean square displacement would be a saturation towards a plateau value
at large time. We also acknowledge that even though what we observe is
in fair agreement with the usual ``power law fluid''
behavior~\cite{desprat05}, the slow initial subdiffusion could be
interpreted as a confinement towards which the bead has not yet
relaxed. Second, we measured the response function by applying optical
tweezers to the bead. The tweezers acted on the bead as a harmonic
potential of spring constant $k_{opt}=
120\text{pN}/\mu\text{m}$. After switching our tweezers on we quickly
translated them by $x_0=0.6 \mu\text{m}$ at time $t=0$, creating a
step force on the bead attached to the cell, and subsequent
deformation of the latter. In the experiment the external force we
applied is hence well approximated by $f_{exp}[x(t),t]=-k_{opt}
(x(t)-x_0) \theta(t)$, assuming the displacement at $t=0$ to be
instantaneous.  The distance between the bead and the center of the
tweezers allowed us to measure the force applied on the bead. By
repeating this experiment we measured the average displacement and the
average force shown in
\fig~\ref{fig:posforce}. The force suddenly increases within a single
time step, i.e. on a time lapse less than $0.004 \text{s}$.  Then, it
slowly decreases towards 0, while the bead displacement increases
following a power law behavior (except in the initial steps of the
relaxation), as shown in the Log-Log scale inset. This is compatible
with the visco-elastic rheology responsible for the sub-diffusivity of
the mean square displacement~\cite{fabry,balland,robert}. Although a
single power law accurately represents the data in the figure, we
chose to fit them by a sum of two power laws, in order to get the best
analytic interpolation required for subsequent analysis.

All the experimental data are discretized in time. We first measure
trajectories of the bead position when no external force is applied.
We have collected $N^0_s=39$ independent sample trajectories of $250
\text{s}$ each with a time step of $0.004 \text{s}$ (each trajectory
consists of $N^0_p=62500$ points). This yields a set of vectors
$x_i^j$, where the subscript $i$ stands for the time position and
superscript $j$ refers to the trajectory number. We compute the mean
square displacement as $\langle{\Delta x^2}\rangle_i=\sum_{j,k} (x_j^k
- x_{j+i}^k)^2/N_s^0 (N_p^0-i)$, from which we take the discrete time
derivative needed in (\ref{eq:fax2}).  A second set of measurements is
taken when we apply the external force on the bead.  Then we measure
both the position $\avg{x}$ and the external force $\avg{f_{exp}}$, as
shown in
\fig~\ref{fig:posforce}. All curves are obtained over a time
window of $25 \text{s}$ with a time step of $\Delta t=0.004 \text{s}$
and ensemble averages are taken over 39 realizations. Our goal is to
recover an estimate of $\avg{F_a(0) x(t)}$ from (\ref{eq:fax2}) by
discretizing integrals and derivatives. To obtain the discrete
response function $R_i$ we had to deconvolve the discrete external
force $\langle{f_{exp}}\rangle$ from $\avg{x}$. For discrete data this
reduces to a matrix inversion, as $\avg{x}_i=\sum_j \Delta
t \langle{f_{exp}}\rangle_{ij} R_j$, where
$\langle{f_{exp}}\rangle_{ij}$ is the triangular matrix with elements
$\langle f_{exp}\rangle((i-j) \Delta t)$. This procedure is generally
sensitive to noise amplification. In order to avoid this phenomenon we
have used {\it ad hoc} numerical fits of the experimental
data. Although the experimental data is fairly well fitted by a single
power law, we have preferred to use a sum of two power laws, which is
more precise at short time. The fit accuracy can be seen
in \fig~\ref{fig:posforce}, where we compare experimental data with
data fit. Likewise, we measured $\gamma_i\equiv \gamma(i \Delta t)$ by
inverting the equation
$\langle{f_{exp}}\rangle_i=\sum_j \left[\frac{d \avg{x}}{dt}
\right]_{ij} \gamma_j$. We
then computed $\sigma_i$ as $1/\beta (\gamma_i + \gamma_{-i})$.

\begin{figure}
\begin{center}
\includegraphics[width=\columnwidth]{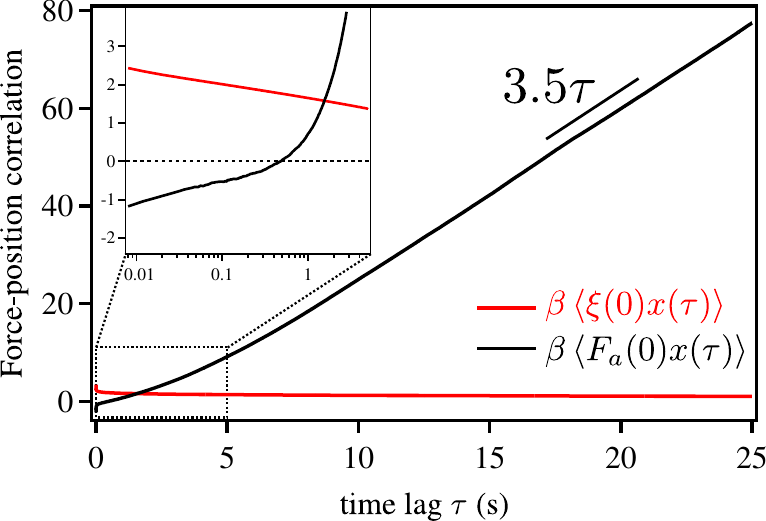}
\end{center}
\caption{Force--position correlation as a function of time lag.
\label{fig:CxF}}
\end{figure}

With the knowledge of $\sigma$, $\Delta x^2(t)$ and $R$ it is then
possible to explicitly compute the cross correlation
function~(\ref{eq:fax2}). We did this by computing integrals as
Riemann sums with step size $\Delta t$ and derivatives as forward
finite differences. The result is shown in \fig~\ref{fig:CxF}, where
we have plotted the correlation functions $\avg{F_a(0) x(t)}$ and
$\avg{\xi(0) x(t)}$. As expected, the thermal noise--position
correlation function is of the order of $\kb T$ and decreases to $\kb
T$ after a time of the order of the second. For comparison, simpler
systems with memory decaying exponentially fast, exhibit for this
function a growth from 0 to $2 \kb T$. Besides, the active
force--position correlation function has a completely different
behavior. First, we observe that the short time behavior exhibits
small yet significant negative correlation. This would mean that on
average the particle moves {\em opposite} to the active force for
about the first half second. At later times, as expected the
correlation function turns positive and grows linearly with time,
which implies that the particle on average moves in the same direction
of the active forces. However we can now quantify this correlation,
which turns out to be of $10-100 \kb T$. Of course we expect that at
later times the correlation function should saturate to a limiting
value. Unfortunately so far our experimentally accessible time window
does not allow to explore this regime (at which one would also see
saturation of the mean square displacement, as discussed earlier).
The figure also shows that for short time ($< 1s$) the effect of
thermal noise prevails, validating our local equilibrium assumption
discussed earlier. At large times the active force dominates, and the
crossover timescale between these two regimes is of the order of 1s.
This seems to corroborate a scenario for which the short time behavior
has equilibrium--like properties, from which the sub-diffusive nature
of visco--elasticity emerges, while the long time behavior is strongly
governed by active non--equilibrium forces. Finally we briefly comment
on the initial anti--correlation exhibited by the active
force--position correlation function. This feature appears for the
first half second of the measurements. Even though in this time range
the data fit (as shown in
\fig~\ref{fig:posforce}) is less accurate than at later times, the
anti-correlation seems to be quite robust~\footnote{We checked that by
  considering overestimations and underestimations of the fitted data
  the anti-correlation is always present.}. So far we do not have a
  clear interpretation for this phenomenon. However, we can get some
  hints from the following scenario. As we have seen, the short time
  dynamics could be seen as an equilibrium--like diffusion in a
  localized potential determined by the local structure of the cell
  cortex. Then, the effect of the active forces at these timescales
  could generate a displacement of this potential. We have checked
  that this simple model (where we are neglecting memory effects) is
  able to exhibit such a negative correlation. We cannot however yet
  reproduce quantitatively our observed results within this simple toy
  model, whose refinements are left for future work.
Another useful information that can be extracted from the correlation
function is concerned with the power spent by the active forces on the
bead. This can be estimated from the time derivative $d/dt \avg{F_a(0)
  x(t)}$ when $t \to 0$. Our experimental data shows that this power
is about $30 \kb T/s$. Assuming a ballistic--like behavior, we
estimate from \fig~\ref{fig:msd} a typical velocity of $v\sim 5 \times
10^{-3} \,\mu\text{m/s}$ (compatible with $0.08 \mu\text{m/s}$
observed in~\cite{finer94}). This leads to an applied force of almost
25 pN, which would mean that a small number (five or less) of motors
are contributing to the bead motion.

The equilibrium fluctuation--dissipation theorem relates the response
of the system to a small perturbation to spontaneous fluctuations in
equilibrium. Out-of-equilibrium extensions involve, besides
dissipative aspects, also kinetic aspects where the dynamical activity
appears within an extra correlation function with the physical
observable under scrutiny. For overdamped Langevin dynamics, and even
for dynamics with strong memory effects, the latter is directly
related to the non--equilibrium forces driving the system out of
equilibrium. Understanding the properties of these forces in the
dynamics of living cells is a problem of its own. With this new
theoretical tool we have shown how to access some of the properties of
these forces. We have been able to quantify the time scale at which
active forces driven fluctuations win over thermal ones. Another
quantitative spin-off is an estimate for the power dissipated by the
active forces into the system, which turns out to be three orders of
magnitude smaller than the chemical power injected into the underlying
motors by the ATPase steps. Another conclusion that we have arrived at
is the strong slaving of the probe motion to the active forces, which
are coherently followed. We have thus demonstrated that extracting
useful information is indeed possible, and this paves the way to a
number of improvements or generalizations. An immediate modification
of the experimental setup would involve the use of an optical trap to
render the unperturbed state stationary, and then pulling on the
particle by displacing the trap. Making the system stationary would
eliminate time drifts and would simplify the subsequent theoretical
analysis by allowing to focus solely on active forces. Other simple
improvements of the present work would, for instance, entail tracking
higher moments of the bead positions. This could be done by attempting
to vary, in the latter stationary setup, the trap's stiffness, instead
of the position of its minimum. Even if the rheological meaning of the
response function would then be lost, other relevant information
concerning fluctuations would then be gained. We leave these projects
for future work.

\acknowledgments We acknowledge several useful discussions with Julien
Tailleur and Damien Robert.

\bibliographystyle{eplbib}
\bibliography{activefdt}

\end{document}